\documentstyle[12pt]{article}
\def\be{\begin{equation}}
\def\ee{\end{equation}}
\def\ba{\begin{array}}
\def\ea{\end{array}}
\def\beqn{\begin{eqnarray}}
\def\eeqn{\end{eqnarray}}
\def\nonum{\nonumber}

\def\vus{$|V_{us}|$}
\def\vcb{$|V_{cb}|$}

\def\vtd{$|V_{td}|~$}

\def\rub{$|\frac {V_{ub}}{V_{cb}}|$}

\begin{document}

\title{UNIQUE MASS TEXTURE FOR QUARKS AND LEPTONS}
\author{Monika Randhawa$^1$, V. Bhatnagar$^1$, P.S. Gill$^{1,2}$ and
 M. Gupta$^1$ \\
$^1$  {\it Department of Physics, Panjab University, Chandigarh-
  160 014, India.}\\
 $^2$  {\it Sri Guru Gobind Singh College, Chandigarh-160 026, India.}}
  \date{11 March 1999}
 \maketitle
  \begin{abstract}
  Texture specific quark mass matrices which are hermitian and
  hierarchical are  examined in detail . In the case of
  texture 6 zeros matrices, out of sixteen possibilities examined
  by us, none is able to fit the
  low energy data (LED),
  for example,
  \vus = 0.2196 $\pm$ 0.0023,
 \vcb = 0.0395 $\pm$ 0.0017,  \rub = 0.08 $\pm$ 0.02,  \vtd 
 lies in the range 0.004 - 0.013 (PDG).
   Similarly none of the 32 texture 5
  zeros mass matrices considered is able to reproduce LED.
  In particular,
  the latest data from LEP regarding \rub $~(=0.093\pm0.016)$ rules
  out all of them.
   In the texture 4 zeros case, we find that there is a unique
  texture structure for $U$ and $D$ mass matrices
  which is able to fit the data.
  \vskip .2cm
  PACS number(s): ~12.15.Ff, 14.60.Pq, 96.60.Jw
\end{abstract}

 The raison d'$\hat{e}$tre for the existence of three
 well seperated families
 of charged fermions remains ununderstood in the context of present
 day high energy physics. The mystery regarding the fermion masses has
 further deepened with the observation of ``neutrino oscillations"  by
 the  Superkamiokande Collaboration \cite{superkam}
 implying nonzero masses
 for the neutrinos and thus, giving for the first time a strong signal
 for the physics beyond the standard model. In the absence of any deeper
  understanding of fermion masses, attempts have been made on the one
  hand to understand arbitrary standard model Yukawa couplings of the
   fermions from more fundamental theories such as GUTs \cite{th1}
  composite models
 \cite{th2}, left right symmetric models \cite{th3}, etc.. On the other
  hand, attempts have been made to discover phenomenological
  quark mass matrices, which are in tune with the
  low energy data(LED).
   In this regard,  specific {\it ansatz} for the quark
   mass matrices have been tried with a fair degree of success \cite{frz}
  -\cite{kang} to explain quark mixing matrix. Similarly
  phenomenological neutrino mass matrices have been considered
  \cite{kang,gillneut}
   which attempt to accomodate simultaneously the
   Solar Neutrino Problem (SNP),
   Atmospheric Neutrino Problem (ANP) and Neutrino
   Oscillations observed by LSND.

  The purpose of the present letter is to find possible textures
  \cite{texdef} for
   mass matrices which are able to accomodate LED regarding
   quark mixing as well as neutrino oscillations required to explain
   SNP, ANP and LSND oscillations.
  To this end, we carry out a
  detailed and exhaustive analysis of a large number of texture specific
  quark  mass matrices (almost 52) and try to find out a set of
  quark mass
  matrices which can accomodate LED. After having restricted the number
  of quark mass matrices, we assume a similar texture structure for
  neutrino mass matrices and study its implications for three neutrino
  anomalies.

  Before one can take an extensive analysis of mass matrices, it should
   be borne in mind that the number of
  free parameters available with the general mass matrices is larger
  than the physical observables. For example,  if no restrictions are
  applied, there are $36$ free parameters to describe $10$ physical
  observables i.e. $6$ quark masses ,
    $3$ mixing angles and one CP violating phase. Therefore to develope
 viable phenomenological quark mass matrices one has to limit the number
 of free parameters in the mass matrices.
 It is, therefore, desirable to invoke certain broad guiding principles
  based on general considerations \cite{peccei96,rrr}
   borne out of experimental data \cite{pdg} and insight gained from
   the past such analyses \cite{albright}-\cite{gill}.
   These guidelines constrain the present analysis
 to manageable number of mass matrices by restricting the number of free
   parameters of general quark mass matrices. In fact, it is also the
 purpose of present letter to consolidate and reiterate these guiding
   principles.

 In this context, we first make use of the polar decomposition theorem
     of matrix algebra, by which, one can always express a
  general mass matrix as a product of a hermitian and a unitary matrix.
  Therefore, without loss of generality, we can consider quark mass
   matrices to be hermitian as the unitary matrix can be absorbed
    in the right handed
  quark fields. This immediately brings
   down the number of free parameters from $36$ to $18$.

   The hierarchical pattern of quark masses as well as
  those of mixing angles immediately suggests that one should start with
  mass matrices whose elements follow hierarchy. 
  This is borne out of several past and
  present analyses \cite{rrr}-\cite{gillneut}.
  One can think of mass matrices whose elements do not exhibit hierarchy,
  nevertheless are still able to reproduce the quark masses, however
  such mass matrices do not satisfy the criterion of ``naturalness"
  proposed recently by Peccei and Wang  \cite{peccei96}. Therefore,
  we assume that the
    elements of hermitian quark mass matrices follow the pattern:
 \be
  M = \left[ \ba {ccc} a_{11} & a_{12} & a_{13} \\
                              a_{21} & a_{22} & a_{23} \\
                                a_{31} & a_{32} & a_{33}. \ea \right],
 \nonum \ee
\begin{center}
$ a_{11} <  |a_{12}| \sim |a_{13}| \ll a_{22} \sim |a_{23|} \ll a_{33} $ 
\end{center}
   The famous Fritzsch {\it ansatz} and subsequent generalizations \cite
   {albright,xing,gupjon,gill} have all considered hierarchical matrices.

  By using polar decomposition theorem, though the number of parameters is
  brought down to $18$,  it is still larger than the number of observables,
   therefore it needs to be brought down further. In this context,
   following  Weinberg \cite{weinberg} and Fritzsch \cite{frz}, the
   strategy has been to assume that mass matrices for fermions have
   certain ``textures"  imposed on them by some underlying symmetries
   or these could
   be purely phenomenological {\it ansatze}. These textures
   allow one to derive some interesting ``predictions" which can then
   be compared with the experiments.  Therefore, in order
   to keep free parameters under control, one therefore starts with
   texture specific {\it ansatze}.

   Before proceeding further, it is perhaps natural to ask whether the
   assumptions of hermiticity, hierarchy
   and textures are preserved when one scales down from GUT scale to
 low energy as the mass matrices are usually derived at the GUT scale.
 This question has been examined in detail \cite{xing,peccei96,rrr}
  and it has been shown that the hierarchical structure of mass
  matrices is not affected,
  whereas the texture structure and hermiticity is broken to a minor
 extent leaving phenomenological consequences unaltered. The texture
 structure, however, is maintained by the RG equations
  if it is ensured by additional symmetry. We, therefore, consider at
    low energy  phenomenological texture specific mass matrices 
    which are hermitian and hierarchical.

A well determined  quark mixing matrix \cite{pdg}, in particular the 
   knowledge of elements $|V_{us}| $, $|V_{cb}| $,
   and  \rub,  leads to the vital clues
     for the possible structures of mass matrices. In this 
context, a survey of some of the past analyses \cite{frz}-\cite{gupjon}
  as well as our own investigations \cite{gill} suggest that $|V_{us}|$
  is given by $\sqrt{m_d/m_s}$ and a very small correction term  
   $\sqrt{m_u/m_c}$ while $|V_{cb}| $ is
   given by $m_s/m_b$ and a correction term 
  $m_c/m_t$. Thus $|V_{us}| $ is controlled largely by D type of quark
   mass matrices, whereas in the case of $|V_{cb}| $ both U and D 
   sectors contribute significantly. These facts can give us vital 
   clues about the positioning of the texture zeros in the
   phenomenological mass matrices.
  For the sake of simplicity, it is desirable that the U and D type of
  the mass matrices be taken to have identical texture
  structures \cite{albright,xing}.

  Keeping in mind above broad guidelines, one could start with the
  specific texture structures in the U type and D type of mass matrices
   \cite{rrr}.
  Before going into the detailed structures of $3 \times 3$ hierarchical
  mass matrices, one would like to note that in a phenomenological
  {\it ansatz}, the $(1,1)$ element can always be
  taken to be equal to zero, because  a nonzero $(1,1)$ element
  leads only to the rescaling of the lightest quark masses
  in both U and D type of mass matrices \cite{albright}. It can be very 
  easily seen that the maximum number of texture zeros which can 
  be considered for U or D type
  of mass matrices have to be three. More than three would lead to at
  least one of the quark masses to be zero. Therefore we start with
  texture 3 
zeros type of matrices.  In order to have non trivial mixings of three
generations as well as keeping in mind guidelines mentioned above only
   following non trivial texture 3 zeros structures are possible:
    
  \begin{center}
  \be
  $$ $M_3$(I)~~~ :~~~ with texture zeros at $(1,1)$, $(1,3)$
  and $(2,3)$ \\
  $ M_3$(II) ~  :~~~ with texture zeros at $(1,1)$, $(1,2)$
  and $(2,3)$ \\
 $M_3$(III) :~~~ with texture zeros at $(1,1)$, $(2,2)$
 and $(1,3) $ \\
  $ M_3$(IV)~  :~~~ with texture zeros at $(1,1)$, $(2,2)$
  and $(2,3)$ $$
 ~~~~~~~~~~~~~~~~~~~~~~~~~\ee
  \end{center}

 \noindent
where $(1,1)$ etc. correspond to zero at the position of first row and
  first column of mass matrix and so on. In general $M_u$ and $M_d$ 
  could be any of the four matrices mentioned above, resulting in 16
  combinations. However, if $M_u$ and $M_d$ are taken to have parallel 
  texture structure, we are left with only four possibilities.
   All these matrices have been diagonalized exactly. The corresponding
 CKM matrices can be found easily \cite{gill}
  and checked against the experimental values of CKM matrix elements,
 for example,  \vus = 0.2196 $\pm$ 0.0023,
 \vcb = 0.0395 $\pm$ 0.0017,  \rub = 0.08 $\pm$ 0.02,  \vtd 
 lies in the range 0.004 - 0.013 \cite{pdg}.
 In table 1, we have summarized the expressions for the CKM matrices 
  corresponding to the four possibilties mentioned above. Without going
  into the details of the methodology for analyzing such CKM matrices,
  we refer the reader to our earlier work \cite{gill}. However,
  we would like to emphasize that the quark masses taken for the CKM
  matrix analysis
  correspond to masses at $1 ~GeV$ \cite{gasleut}, 
  for example, $m_u=0.0051\pm0.0015~GeV,$ $m_d=0.0089\pm0.0026~GeV$,
   $m_s=0.175\pm0.055~GeV$, $m_c=1.35\pm0.05~GeV$,
   $m_b=5.3\pm0.1~GeV$  and $m_t=300\pm50~GeV$.

 \pagebreak
   \begin{table}
   \begin{tabular}{|c|c|c|c|c|} \hline
    & $M_i$  & $V_{us}$ & $V_{cb}$ & $V_{ub}$ \\  \hline
          I & $\left( \ba  {ccc}  0 & A_i & 0 \\ A_i^* & D_i & 0 \\
            0 & 0 & C_i \ea \right)$ 
          & $c - ae^{i\phi}$ & $0$ & $0$ \\
  \hline
  II & $\left(  \ba  {ccc} 0 & 0 & A_i \\ 0 & D_i & 0\\
             A_i^* & 0 & C_i \ea \right)$ 
          & $0$ & $0$ & $-cde^{i\phi} + ab$ \\ 
  \hline
  III & $\left( \ba  {ccc} 0 & A_i & |B_i| \\ A_i^* & 0 & 0 \\
           |B_i| & 0 & C_i \ea \right)$ &
              $\ba {l} a - ce^{i\phi}\\  + abd \ea $ 
  & $\ba {l} -d + acd^3e^{i\phi} \\ + b \ea $
   & $\ba {l}  -ad - cd^3e^{i\phi} \\ + ab \ea $
  \\ \hline
  IV & $\left( \ba  {ccc} 0 & A_i & 0 \\ A_i^* & 0 & |B_i| \\
            0 & |B_i| & C_i \ea \right)$
             & $\ba {l} -ce^{i\phi} + a \\
            + abd \ea$ 
 & $ \ba {l} -acd^3e^{i\phi} + d \\ -b \ea $
& $ \ba {l} cd^3e^{i\phi} +ad \\
 -ab \ea $ \\ \hline \hline
\end{tabular}
\caption{Expressions for CKM matrix elements $V_{us}$, $V_{cb}$ and
  $V_{ub}$ corresponding to $M_i$ ($i=u,d$) listed in column I. The
  symbols used in the table are:
   $a=\sqrt{m_u/m_c}$, $b=\sqrt{m_c/m_t}$,
 $c=\sqrt{m_d/m_s}$,
 $d=\sqrt{m_s/m_b}$, $A_i=|A_i|e^{i\alpha_i}$,
 $\phi=\alpha_u-\alpha_d.$}
\end{table}
\vskip .2cm
From the table 1, it is clear that possibilities I and II are
completely ruled out. In the case of III possibility \vcb ~cannot
be fitted even after full variation of input parameters,
as can be checked from the
expression given in the table. The possibility IV is the famous Fritzsch
 {\it ansatz}, which again gives
$|V_{cb}| $ much above the present experimental value, as has also been
   pointed out in a large number of earlier analyses.
 For the sake of completeness, we have also carried out investigations
 of the possibilities where $M_u$ and $M_d$ don't have parallel
 structures. Twelve such possibilities are there, which  are not
 listed here. We find that
  none of the structures is viable. It may be noted that in texture 
  6  zeros, Fritzsch {\it ansatz}
  is perhaps the best and that sets the tone for future modifications. 

 A strict adherence to parallel texture structure for $M_u$  and $M_d$
  type of mass matrices rules out texture 5 zeros mass matrices,
  however, such matrices have been discussed in literature.
 Therefore, we have also
 included here a discussion of such matrices. Texture 5 zeros 
    mass matrices would have either of the two, $M_u$ or $M_d$,
    being of texture 2 zeros type. As texture 3 zeros
  matrices have already been listed, therefore we list all the possible
  texture 2 zeros mass matrices compatible with the 
  guidelines enunciated above.
\pagebreak
 \begin{center}
 \be
$$ $ M_2$(I)~~~ :~~~ with zeros at $(1,1$)  and $(1,2) $ \\
$ M_2$(II) ~  :~~~ with zeros at $(1,1)$ and $(1,3) $ \\
 $M_2$(III) :~~~ with zeros at $(1,1)$ and $(2,3) $ \\
$ M_2$(IV)~  :~~~ with zeros at $(1,1)$ and $(2,2)$     $$
~~~~~~~~~~~~~~~~~\ee
\label{4 zeros}
\end{center}
 All these matrices are exactly diagonalizable, except for $M_2(IV)$
 where diagonalization can be achieved perturbatively \cite{rasin}.
  To obtain texture 5 zeros mass matrices, one has to combine any of
  the matrices given in eqn. 2 with any
of the matrices in eqn. 3. This leads to $32$ possibilities for the
 texture 5 zeros mass matrices. These possibilities include the five
 such examples discussed by RRR \cite{rrr}. The examples
 considered by RRR are
 ruled out by the present data \cite{pdg}, particularly the recent
 measurement of \rub at LEP \cite{lep} rules these out unambiguously.
 In the table 2, we present some of the examples of $M_u$ and $M_d$
  constituting texture 5 zeros matrices and not considered by RRR.
 \pagebreak
 {\scriptsize
   \begin{table}
        \begin{tabular}{|c|c|c|c|c|c|} \hline
   & & & & & \\ & $M_u$ & $M_d$ & $V_{us}$ & $V_{cb}$ & $V_{ub}$ \\
     & & & & & \\ \hline & & & & & \\
  I & $\left( \ba  {ccc}  0 & A_u & 0 \\ A_u^* &  D_u & |B_u|  \\
            0 & |B_u| & C_u \ea \right)$ &
  $ \left( \ba  {ccc} 0 & A_d & 0 \\ A_d^* & 0 & |B_d|  \\
  0 & |B_d| & C_d \ea \right)$ & $\ba {l} -ce^{i\phi}
  + aR_t^{'\frac{1}{2}}\\
  +adR_t^{"\frac{1}{2}} \ea$ & $\ba {l} -acd^3e^{i\phi}
  +dR_t^{'\frac{1}{2}}   \\
  -R_t^{"\frac{1}{2}} \ea $ & $\ba {l} cd^3e^{i\phi} +adR_t
  ^{'\frac{1}{2}} \\ -aR_t^{"\frac{1}{2}} \ea $ \\
             & & & & &\\ \hline
 & & & & &\\ II & $\left(  \ba  {ccc} 0 & |A_u| & B_u\\ |A_u| & 0 & 0\\
             B_u^* & 0 & C_u \ea \right)$ &
$\left(  \ba  {ccc} 0 & |A_d| & 0 \\ |A_d| & D_d & B_d \\
  0 & B_d^* & C_d \ea \right)$ & $\ba {l} ca- R_b^{'\frac{1}{2}}\\
  -abR_b^{"\frac{1}{2}}e^{i\phi}\ea $ & $\ba {l} -cd^2(\frac{R_b^{"}}
  {R_b^{'}})^{\frac{1}{2}}
\\ +aR_b^{"\frac{1}{2}}\\ +bR_b^{'\frac{1}{2}}e^{i\phi}\ea $ &
$\ba {l} -acd^2(\frac{R_b^{"}}{R_b^{'}})^{\frac{1}{2}}\\
-R_b^{"\frac{1}{2}}
  \\+abR_b^{'\frac{1}{2}}e^{i\phi}\ea $ \\ & & & & &\\
  \hline
 & & & & &\\ III & $\left( \ba  {ccc} 0 & 0 & A_u \\ 0 & D_u & |B_u| \\
            A_u^* &  |B_u|  & C_u \ea \right)$ &
$ \left( \ba  {ccc} 0 & A_d & 0 \\ A_d^* & 0 & |B_d| \\
           0 & |B_d| & C_d \ea \right)$ &
$\ba {l} c(1-\frac{(ab)^2}{R_t})^{\frac{1}{2}}e^{i\phi}\\
+a(\frac{R_t^{"}R_t^{'}}{R_t})^{\frac{1}{2}}\\
 +ad(R_t-(ab)^2)^{\frac{1}{2}} \ea $
  & $\ba {l} acd^3(\frac{R_t^{"}}{R_t})^{\frac{1}{2}}e^{i\phi}\\
  + d(\frac{R_t^{'}(R_t-(ab)^2)}{R_t})^{\frac{1}{2}}  \\
  -R_t^{"\frac{1}{2}} \ea $
   & $\ba {l}  -cd^3(1-\frac{(ab)^2}{R_t})^{\frac{1}{2}}e^{i\phi}\\
   + ad(\frac{R_t^{"}R_t^{'}}{R_t})^{\frac{1}{2}}  \\
    - a(R_t-(ab)^2)^{\frac{1}{2}} \ea $
   \\& & & & & \\ \hline
& & & & &\\ IV & $\left( \ba  {ccc} 0 & A_u & 0 \\ A_u^* & D_u & 0\\
                  0 & 0 & C_u \ea \right)$  &
$ \left( \ba  {ccc} 0 & 0 & A_d \\  0 & D_d & |B_d| \\
           A_d^*  & |B_d| & C_d \ea \right)$
 & $\ba {l} +c(\frac{R_b^{"}}{R_b})^{\frac{1}{2}}e^{i\phi}\\
  + a(\frac{R_b^{'}(R_b-(cd)^2)}{R_b})^{\frac{1}{2}} \ea $ 
 & $ \ba {l} acd^2(\frac{R_b^{'}}{R_b})^{\frac{1}{2}}e^{i\phi}\\
  -(\frac{R_b^{"}(R_b-(cd)^2)}{R_b})^{\frac{1}{2}} \ea $
& $ \ba {l} cd^2(\frac{R_b^{'}}{R_b})^{\frac{1}{2}}e^{i\phi}  \\
 + a(\frac{R_b^{"}(R_b-(cd)^2)}{R_b})^{\frac{1}{2}} \ea$  \\
  & & & & & \\ \hline
& & & & &\\ V & $\left( \ba  {ccc} 0 & 0 & |A_u|  \\ 0 & D_u & B_u\\
                  |A_u| & B_u^* & C_u \ea \right)$  &
$ \left( \ba  {ccc} 0 &  A_d & |B_d| \\ A_d^* & 0 & 0 \\
            |B_d| & 0 & C_d \ea \right)$ &
      $\ba {l} (1-\frac{(ab)^2}{R_t})^{\frac{1}{2}}\\
       + ac(\frac{R_t^{"}R_t^{'}}{R_t})^{\frac{1}{2}}e^{i\phi} \\
         -ad(R_t-(ab)^2)^{\frac{1}{2}}  \ea $
  & $\ba {l} +ad(\frac{R_t^{"}}{R_t})^{\frac{1}{2}}\\
   + cd^3(\frac{R_t^{'}(R_t-(ab)^2)}{R_t}) ^{\frac{1}{2}}e^{i\phi} \\
    - R_t^{"} \ea $
   & $\ba {l}  -d(\frac{R_t-(ab)^2}{R_t})^{\frac{1}{2}}\\
    +acd^3(\frac{R_t^{"}R_t^{'}}{R_t})^{\frac{1}{2}}e^{i\phi} \\
     - a(R_t-(ab)^2)^{\frac{1}{2}} \ea $
   \\& & & & & \\ \hline \hline
\end{tabular}
  \caption{Expressions for CKM matrix elements $V_{us}$, $V_{cb}$ and
  $V_{ub}$ corresponding to $M_u$ and $M_d$ listed in column I and II
  respectively. The symbols used are as defined for table 1 and
   $B_i=|B_i|e^{i\alpha_i}$, $R_t=D_u/m_t$,  $R_t^{'}=1-R_t$,
   $R_t^{"}=b^2+R_t$,
  $R_b=D_d/m_b$,  $R_b^{'}=1-R_b$, $R_b^{"}=d^2+R_b$.}
 \end{table} }

 Proceeding in the same manner as that of texture 6 zeros matrices, we 
 find that all the cases considered in table 2 are ruled out. In
 particular the possibility V is ruled out as \vus ~can not be reproduced. 
  In the cases I, III and
   IV, \vus ~can be reproduced but \vcb ~can not be fitted. 
     In the case of possibility II, \vus ~can be
 reproduced for large values of $R_b$, however that makes \vcb ~too
  large to fit the data.
  Similarly, we have exactly derived the CKM matrix for
   the rest of the texture 5 zeros possibilities and found
   that none of these is able to reproduce full CKM matrix, even
   after full variation of all the parameters.

  After having ruled out the texture 5 zeros mass matrices, it is
   natural to consider texture 4 zeros matrices. In table 3, we have
   listed texture 4 zeros mass matrices, where $M_u$ and $M_d$ are
   respectively 2 zeros type and have parallel structures. 
 \pagebreak
  {\tiny
   \begin{table}
   \begin{tabular}{|c|c|c|c|c|} \hline
   & & & & \\ & $M_i$  & $V_{us}$ & $V_{cb}$ & $V_{ub}$ \\ & & & & \\
   \hline
  & & & & \\
       I & $\left( \ba  {ccc}  0 & 0 & A_i  \\ 0 &  D_i & |B_i|  \\
           A_i^* & |B_i| & C_i \ea \right)$ &
 $\ba {l} -c((1-\frac{(ab)^2}{R_t})\frac{R_b^{"}}{R_b})
 ^{\frac{1}{2}}e^{i\phi}\\
 + (\frac{R_t^{"}R_t^{'}R_b^{'}}{R_t}
 (1-\frac{(cd)^2}{R_b}))^{\frac{1}{2}}\\
  + a( R_b^{"}(R_t-(ab)^2))^{\frac{1}{2}} \ea $ &
$\ba {l} acd^2(\frac{R_t^{"}R_b^{'}}{R_tR_b})^{\frac{1}{2}}e^{i\phi}\\
 +(\frac{(R_t-(ab)^2)R_t^{'}(R_b-(cd)^2)R_b^{"}}
 {R_bR_t})^{\frac{1}{2}} \\
   -(R_b^{'}R_t^{"})^{\frac{1}{2}} \ea $ &
  $\ba {l} -cd^2(\frac{(R_t-(ab)^2)R_b^{'}}
  {R_tR_b})^{\frac{1}{2}}e^{i\phi}\\
  +(\frac{R_t^{"}R_t^{'}(R_b-(cd)^2)R_b^{"}}{R_bR_t})^{\frac{1}{2}} \\
  -a(R_b^{'}(R_t-(ab)^2))^{\frac{1}{2}} \ea $ \\
 & & & &\\ \hline
 & & & &\\ II & $\left(  \ba  {ccc} 0 & A_i & |B_i|\\ A_i^* & D_i & 0 \\
              |B_i| & 0 & C_i \ea \right)$ &
   $\ba {l} +\frac{1}{bd}(((ab)^2-R_t)R_t^{'}
   R_b^{'}R_b^{"})^{\frac{1}{2}}\\
  -\frac{R_t^{'}R_b^{'}}{bd}(\frac{R_t^{"}(R_b-(cd)^2)}
  {\Delta_t\delta_b})^{\frac{1}{2}}e^{i\phi}\\
  +\frac{R_b^{"}}{bd}(\frac{R_t^{'}R_b^{'}R_t^{"}(R_t-(ab)^2)}{\Delta_t
  \Delta_b})^{\frac{1}{2}} \ea $ &
   $\ba {l} -\frac{1}{b}(R_t^{"}R_t^{'}R_b^{'}R_b^{"})^{\frac{1}{2}}\\
+\frac{R_t^{'}R_b^{"}}{b}(\frac{(R_t-(ab)^2)(R_b-(cd)^2)}
{\Delta_t\Delta_b})
  ^{\frac{1}{2}}e^{i\phi}\\
  +\frac{R_t^{"}R_b^{'}}{b}(\frac{R_t^{'}R_b^{'}}{\Delta_t
  \Delta_b})^{\frac{1}{2}} \ea $ &
   $\ba {l} -\frac{1}{b}(((ab)^2-R_t)R_t^{'}R_b^{'}R_b^{"})
   ^{\frac{1}{2}}\\
-\frac{R_t^{'}R_b^{"}}{b}(\frac{R_t^{"}((cd)^2-R_b)}{\Delta_t\Delta_b})
  ^{\frac{1}{2}}e^{i\phi}\\
  +\frac{R_b^{'}}{b}(\frac{R_t^{'}R_b^{'}R_t^{"}((ab)^2-R_t)}{\Delta_t
  \Delta_b})^{\frac{1}{2}} \ea $  \\ & & & &\\
  \hline
  & & & &\\ III & $\left(  \ba  {ccc} 0 & A_i & B_i\\ A_i^* & 0 & D_i\\
             B_i^* & D_i & C_i \ea \right)$ &
   $\ba {l} ce^{i\phi}-a\\
  +abd \ea $ &
   $\ba {l} 2acde^{i\phi} + d\\
   - b \ea $ &
   $\ba {l} 2cde^{i\phi} -ad\\
  - ab  \ea $  \\ & & & &\\
  \hline
 & & & &\\ IV & $\left(  \ba  {ccc} 0 & A_i & 0\\ A_i^* & D_i & |B_i|\\
            0 & |B_i| & C_i \ea \right)$ &
   $\ba {l} -ce^{i\phi} + a(R_t^{'}R_b^{'})^{\frac{1}{2}}\\
  + a(R_b^{"}R_t^{"})^{\frac{1}{2}} \ea $ &
   $\ba {l} -acd^2(\frac{R_b^{"}}{R_b^{'}})^{\frac{1}{2}}e^{i\phi}\\
  + (R_t^{'}R_b^{"})^{\frac{1}{2}}\\
  - (R_t^{"}R_b^{'})^{\frac{1}{2}} \ea $ &
   $\ba {l} cd^2(\frac{R_b^{"}}{R_b^{'}})^{\frac{1}{2}}e^{i\phi}\\
  + a(R_t^{'}R_b^{"})^{\frac{1}{2}}\\
  - a(R_b^{'}R_t^{"})^{\frac{1}{2}}
  \ea $  \\
  & & & & \\
  \hline \hline
\end{tabular}
  \caption{Expressions for CKM matrix elements $V_{us}$, $V_{cb}$ and
  $V_{ub}$ corresponding to $M_i$ ($i=u,d$) listed in column I. The
  symbols used are as defined for tables 1 and 2 and
$\Delta_b=1-2R_b,$ $ \Delta_t=1-2R_t $.}   
\end{table}}
  
 Following the procedure outlined earlier, one
can easily see that possibilities I, II and III are not able to
reproduce \vus, \vcb ~and \rub ~simultaneously, therefore we are left
with only the possibility IV.  In table 4, we have presented
 the numerical values of \vus, \vcb, \rub ~and \vtd ~as a function
  of independent parameters $R_t$ and $R_b$. To limit the number of
  possibilities we have considered only those cases where $R_b=R_t.$

 \pagebreak
\begin{center}
\begin{table}
  \begin{tabular}{|c|c|c|c|c|c|} \hline
   & & & & & \\  $R_t$ & $R_b$ & \vus & \vcb & \rub & \vtd \\
    & & & & & \\ \hline
  & & & & & \\ 0.075 & 0.075 & 0.22 & 0.044 & 0.0829 &  0.009 \\
  & & & & & \\  \hline
& & & & & \\     0.080 & 0.080 & 0.22 & 0.043 & 0.085 & 0.009 \\
& & & & & \\  \hline
 & & & & & \\    0.085 & 0.085 & 0.22 & 0.042 & 0.087 & 0.009 \\
 & & & & & \\  \hline
& & & & & \\     0.090 & 0.090 & 0.22 & 0.041 & 0.089 & 0.009 \\
& & & & & \\  \hline
& & & & & \\     0.950 & 0.950 & 0.22 & 0.040 & 0.091 &  0.008 \\
& & & & & \\  \hline
& & & & & \\     0.100 & 0.100 & 0.22 & 0.039 & 0.094 & 0.008 \\
& & & & & \\  \hline
& & & & & \\     0.105 & 0.105 & 0.22 & 0.038 & 0.096 & 0.008 \\
& & & & & \\  \hline
& & & & & \\     0.110 & 0.110 & 0.22 & 0.037 & 0.099 & 0.008 \\
& & & & & \\  \hline
    \end{tabular}
\caption{Calculated values of $|V_{us}|$,  $|V_{cb}|$, $\frac{|V_{ub}|}
{|V_{cb}|}$ and  $|V_{td}|$ corresponding to $M_u$ and $M_d$ of IV row
 in table 3.}
    \end{table}
  \end{center}
     A quick look at the table 4 shows that the calculated values of
     \vus, ~\vcb, and \vtd ~are well within the experimental bounds
     \cite{pdg}. The details of the CKM phenomenology with
     CKM matrix derived from  texture 4 zeros mass matrices will be
     discussed elsewhere.

   After having found a viable texture structure for quark mass
   matrices, it is natural to ask whether a similar texture could be
   used for neutrino mass matrices or not. In particular, one would
   like to examine whether
   such a texture can generate the required appearance and disappearance
   probabilities for explaining the three neutrino anomalies. In this
   context, recently an analysis has been carried out by Barenboim and
   Scheck \cite{scheck}. In particular, they find a
    mixing matrix, which is able to provide a simultaneous fit to the
    SNP, ANP and LSND oscillations, for example,
 \vskip .2cm
 $ \left[ \ba {ccc} 0.793 & 0.566 & 0.226\\ -0.601 & 0.662 & 0.447\\
        0.103 & -0.490 & 0.865 \ea \right].$     \\
    In our analysis based on texture 4 zeros mass matrices,
    the above matrix can be reproduced by considering the following
    eigenvalues of neutrino masses:
 %$  \left[ \ba {ccc} 0.801 & 0.540 & 0.254\\ 0.595 & 0.713 & 0.372\\
 %       0.067 & 0.448 & 0.892 \ea \right] $ \\
  %\noindent
  $m_1=0.53\times10^{-3}eV$, $m_2=0.1\times10^{-1}eV$ and $m_3=0.5eV.$
Without going into the details (to be discussed elsewhere), we would
   like to emphasize that hierarchical neutrino masses can generate
   the required mixing matrix within the texture four zero
   scenario.

In conclusion, we would like to mention that an extensive analysis of
    a large number of texture specific quark mass matrices, which are
     hermitian and hierarchical has been carried out. Interestingly,
     there are no quark mass matrices with texture 6 zeros and texture
     5 zeros which can fit LED. In the case of texture 4 zeros, there
      is a unique texture with parallel texture structure for $M_u$
      and $M_d$ which fits the data.
 When a similar texture structure is assumed for neutrinos, we are
  able to reproduce a mixing matrix which can accomodate Solar Neutrino
  Problem, Atmospheric Neutrino Problem and the oscillations observed
  at LSND.

  \vskip .2cm
  {\bf ACKNOWLEDGMENTS}\\
 M.R. would like to thank CSIR, Govt. of India, for  the fellowship.
M.R. and P.S.G. would like to thank the chairman, Department of Physics,
for providing facilities to work in the department. P.S.G. acknowledges
the financial support recieved for his UGC project. M.G. would like to
thank C.S. Aulakh and S.D. Rindhani for useful dicussions.

\end{document}